\documentclass[a4paper]{jpconf}
\usepackage{graphicx}
\usepackage[latin1]{inputenc}

\begin{document}

\title{Renormalization group calculation of the uniform susceptibilities in low-dimensional systems}
\author{Hermann Freire, Eberth Corrêa, and Álvaro Ferraz}
\address{Centro Internacional de Física da Matéria Condensada,
Universidade de Brasília, Brasília-DF, C.P. 04513, Brazil}
\ead{hermann@iccmp.org}

\begin{abstract}
We analyze the one-dimensional (1D) and the two-dimensional (2D)
repulsive Hubbard models (HM) for densities slightly away from
half-filling through the behavior of two central quantities of a
system: the uniform charge and spin susceptibilities. We point out
that a consistent renormalization group treatment of them can only
be achieved within a two-loop approach or beyond. In the 1D HM, we
show that this scheme reproduces correctly the metallic behavior
given by the well-known Luttinger liquid fixed-point result. Then,
we use the same approach to deal with the more complicated 2D HM. In
this case, we are able to show that both uniform susceptibilities
become suppressed for moderate interaction parameters as one take
the system towards the Fermi surface. Therefore, this result adds
further support to the interpretation that those systems are in fact
insulating spin liquids. Later, we perform the same calculations in
2D using the conventional random phase approximation, and establish
clearly a comparison between the two schemes.
\end{abstract}

\section{Introduction}

\vspace {0.5cm}

The anomalous electronic properties of the metallic phase of the
copper-oxide high-Tc superconductors are widely believed to be
linked to the manifestation of a non-Fermi liquid state arising from
the two-dimensional (2D) character of those compounds. Motivated by
this, physicists have then been trying to understand theoretically
what are the precise conditions in which Landau's Fermi liquid
paradigm \cite{Landau,Nozieres} (i.e. the scheme based on the
concept of low-lying electronic quasiparticles in the system) breaks
down in 2D models.

There are a number of approaches for accomplishing that task in 2D
systems. The most direct route is through Landau's theory of
spontaneous breaking of a continuous symmetry, often induced by an
instability with respect to a relevant external perturbation. That
case usually leads to a phase transition scenario accompanied by the
realization of a long-range ordered state in the system. Moreover,
it implies the onset of gapless low-energy elementary excitations:
the so-called Goldstone modes (see Goldstone's theorem in Refs.
\cite{Nambu,Goldstone}). This theory provides the framework for
understanding, among several other examples, the long-range magnetic
order realized in the Mott insulating state of the 2D Hubbard model
(HM) exactly at half-filling, and the resulting emergence of gapless
antiferromagnetic magnon excitations, which manifest themselves
solely through the spontaneous breaking of the spin rotation
symmetry in the system.

\begin{figure}[b]
  \includegraphics[width=2.85in]{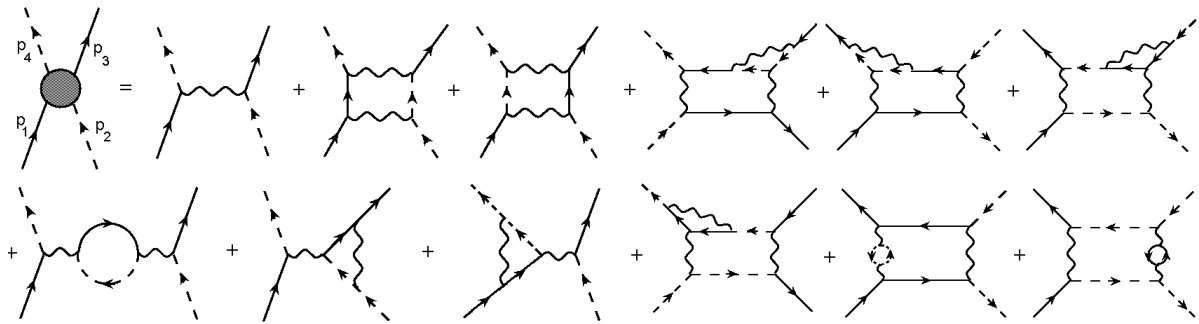}\\
  \caption{The Feynman diagrams for the one-particle irreducible four-point function
  $\Gamma^{(4)}(p_{1},p_{2},p_{3},p_{4})$ in the backscattering
  channel up to two-loop order (third order). The free propagators are represented by either
  solid or dashed lines according to their association with the corresponding branches.}\label{fig2}
\end{figure}

Alternatively, another viable approach for that is produced
quantum-mechanically by the disordering induced by strong
fluctuations resulting from low-dimensionality effects themselves or
by frustration effects in 2D models. These interacting regimes in
turn do not have any broken symmetry realized in the ground state
and, as a result, all the correlations are purely short-ranged in
real space. In view of this, in principle, there can be no gapless
excitations in those cases, since Goldstone's theorem does not apply
anymore. These exotic states cannot be described by the conventional
Landau's theory mentioned before, and are commonly referred to as
spin liquid ground states. Their existence was first predicted by
Anderson long ago in the context of a Heisenberg model defined on a
2D triangular lattice \cite{Anderson1}, and in \cite{Anderson2} a
decade later, and they might possibly hold the key for the
appropriate theoretical explanation of the underlying mechanism of
the high-Tc superconductors.

To understand these issues better, theorists have been generally
looking for similar physics in simpler models. For instance, in
strictly one-dimensional (1D) systems, the situation is far more
understood in that respect \cite{Solyom,Haldane,Voit}. It is by now
very well-established that Landau's Fermi liquid description breaks
down completely in those particular cases. In fact, the appropriate
effective low-energy theory for the 1D metallic state is instead the
so-called Luttinger liquid. This happens due to specific features
associated with 1D, i.e, the always nested Fermi surface (FS)
consisting of only two discrete points ($+k_{F}$ and $-k_{F}$), and
the very restricted phase space available for the elementary
excitations above the corresponding ground state. Consequently, the
low-energy excitation spectrum becomes fundamentally dominated by
bosonic collective excitations given by two types: the holons and
the spinons excitations. These particles in turn imply an exact
separation of the charge and spin degrees of freedom in the system.
As a result, they are most effective in destroying completely the
fundamental picture of well-defined low-lying quasiparticles, which
are the basis of Fermi liquid theory.

In this paper, it is our main goal to investigate the low-energy
properties of the well-known 2D repulsive HM for densities slightly
away from half-filling. To do this, we analyze two central
quantities of the model: the uniform charge and spin
susceptibilities. We choose these two physical quantities in view of
the fact that they naturally provide important information about the
existence or not of low-lying excitations in the system. As will
become clear soon, to perform a full renormalization group (RG)
calculation of those quantities, it is essential to set up at least
a two-loop order RG calculation for it. As a testing ground of this
approach, we first calculate these quantities in the 1D repulsive HM
away from half-filling. In this case, we show that both quantities
approach finite values, and the resulting low-energy description is
indeed a metallic phase (in fact a Luttinger liquid) in agreement
with well-known exact results (Bethe ansatz, bosonization, etc).
Later, we apply the same considerations to the more complicated 2D
HM, which is the case we are mainly interested here. We point out
that there is an interaction regime, where both quantities become
strongly suppressed as one approaches the low-energy limit, thus
adding further support to the interpretation that the resulting
effective theory should be given by an insulating spin liquid state
\cite{spinliquid}. Then, we compare these results with the
conventional random phase approximation (RPA) case, and show that
this latter approximation is not able to detect clearly such an
exotic ground state in the system. Finally, we discuss the physical
origins of the discrepancies of those two results.

\vspace {0.5cm}

\section{Testing ground: The 1D repulsive Hubbard model}

\vspace {0.5cm}

First of all, we define the Hubbard model (HM) defined on an 1D
lattice here. This model describes a system of tight-binding
electrons hopping only to nearest neighbor sites, and interacting
mutually through a local (in real space) repulsive interaction
parametrized by $U>0$. The corresponding Hamiltonian given in
momentum space is

\begin{equation}
H=\sum_{k,\sigma}\xi_{k}\psi^{\dagger}_{k\sigma}\psi_{k\sigma}+\left(\frac{U}{N_{sites}}\right)
\sum_{p,k,q}\psi^{\dagger}_{p+k-q\uparrow}\psi^{\dagger}_{q\downarrow}
\psi_{k\downarrow}\psi_{p\uparrow},
\end{equation}

\noindent where the single-particle energy dispersion is simply
$\xi_{k}=-2t\cos ka-\mu$, and $\psi^{\dagger}_{k\sigma}$ and
$\psi_{k\sigma}$ are the usual creation and annihilation operators
of electrons with momentum $k$ and spin projection
${\sigma}=\uparrow,\downarrow$. Besides,  $\mu$ stands for the
chemical potential, $a$ is the lattice spacing, and $N_{sites}$ is
the total number of lattice sites. Another important parameter here
is the width of the noninteracting band, which is given by $W=4t$.

The FS of the noninteracting system is given by two discrete points
($+k_{F}$ and $-k_{F}$). Since we are mostly interested in the
low-energy properties of this model, we linearize the energy
dispersion about the two Fermi points, i.e.,
$\xi_{k}=v_{F}\left(\left| k\right|-k_{F}\right)$, where
$v_{F}=(\partial \xi_{k}/\partial k)|_{k=k_{F}}$ is the Fermi
velocity of the system. We restrict the allowed single-particle
states to lie within the interval $k_{F}-\lambda \leq
\left|k\right|\leq k_{F}+\lambda$, where $\lambda$ is a fixed
ultraviolet (UV) microscopic momentum cut-off, and $\Omega$ is the
corresponding energy cutoff given by $\Omega=2v_{F}\lambda$. As a
result, we are able to identify two different regions in k-space:
the ``+'' branch linearized about $k=+k_{F}$, and the ``--'' branch
linearized about $k=-k_{F}$. Thus, the corresponding free
propagators may be represented by two different lines in the Feynman
diagrams: solid line for the former, and dashed line for the latter.
In addition, regarding the interaction term, we use the standard
g-ology parametrization \cite{Solyom}, namely, the interaction
processes involving large momentum transfers (near $2k_{F}$) are
described by the $g_{1}$ coupling, whereas the processes involving
small momentum transfers between particles located at different
branches are described by the $g_{2}$ coupling. The former
interactions are usually called backscattering processes, while the
latter are referred to as forward scattering processes. In what
follows, we will deliberately neglect the contributions due to the
so-called Umklapp processes, which are described by the coupling
$g_{3}$, in view of the fact that we will not consider here the
exceptional case of an exact half-filled band in the model.
Moreover, we will make one further approximation and neglect forward
scattering pocesses between particles belonging to the same branch,
which are conventionally described by the coupling constant $g_{4}$.

\begin{figure}[t]
  \centering
  \includegraphics[width=4.1in]{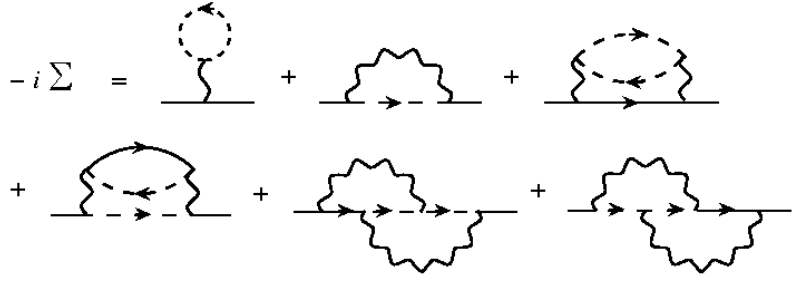}\\
  \caption{The self-energy diagrams up to two-loop order. The
  one-loop order (first order) diagrams are the nonsingular Hartree-Fock terms. The first nonanalyticity
  comes from the two-loop order (second order) diagrams.}\label{fig2}
\end{figure}

In the following, we will use the so-called field-theoretical RG
strategy to tackle this problem. In view of this, it is more
convenient to work here with the Lagrangian of the model instead of
the Hamiltonian. Thus, we write down the Lagrangian of the 1D HM as

\begin{eqnarray}
&&L=\sum_{k,\sigma,a=\pm}\psi_{(a)\sigma}^{\dagger}\left(k,t\right)
\left[i\partial_{t}-v_{F}\left(\left|k\right|-k_{F}\right)\right]\psi_{(a)\sigma}
\left(k,t\right)\nonumber \\
&&-\frac{1}{V}\sum_{p,q,k}
\sum_{\alpha,\beta,\delta,\gamma}\left[g_{2}\delta_{\alpha\delta}\delta_{\beta\gamma}-g_{1}\delta_{\alpha\gamma}\delta_{\beta\delta}\right]\nonumber
\psi_{\left(+\right)\delta}^{\dagger}\left(p+q-k,t\right)\psi_{\left(-\right)\gamma}^{\dagger}\left(k,t\right)
\psi_{\left(-\right)\beta}\left(q,t\right)\psi_{\left(+\right)\alpha}\left(p,t\right),
\\ \label{lagrangian}
\end{eqnarray}

\noindent where $\psi^{\dagger}_{(\pm)}$ and $\psi_{(\pm)}$ are now
fermionic fields associated to electrons located at the $\pm$
branches. The summation over momenta must be appropriately
understood as $\sum_{p}=V/(2\pi)\int dp$ in the thermodynamic limit,
where $V$ is the ``volume'' of the 1D system. Since this should
represent the 1D HM, the Lagrangian is written in a manifestly
$SU(2)$ invariant form and, in addition to this, the couplings are
initially defined as $g_{1}=g_{2}=(V/4N_{sites})U$.

Since the HM is microscopic, the couplings and the fermionic fields
in the Lagrangian are defined at a scale of a few lattice spacings
in real space (i.e at the UV cutoff scale in momentum space). These
parameters are usually inaccessible to every day experiments, since
the latter probe only the low-energy dynamics of the system. In RG
theory, these unobserved quantities are known as the bare
parameters. In fact, when one tries to construct naive perturbative
calculations with such quantities, one is immediately confronted
with several Feynman diagrams, which turn out to be singular in the
low-energy limit. These are the so-called infrared (IR) divergences
in field theory. They appear in both backscattering and forward
scattering four-point vertex corrections (see, for instance, Fig.
1), and in the calculation of the self-energy alike (Fig. 2). Those
divergences do not imply necessarily that perturbation theory is
simply unable to give any reliable prediction of the low-energy
properties of the model, but only that the way it is formulated is
not the appropriate one for this case. Therefore, one must
generalize the perturbative approach so as to eliminate such
singularities in the calculations. This is the strategy of the
so-called renormalized perturbation theory, i.e, one rewrites all
unobserved bare parameters in terms of the associated renormalized
(or observable) quantities. The difference between them will be
given by the so-called counterterms, whose essential role is to
absorb all divergences up to a given order in perturbation theory.
If this program is successfully accomplished, then the theory is
said to be properly renormalized.

Following this, we have

\vspace {-0.3cm}

\begin{eqnarray}
\psi_{(a)\sigma}&&\rightarrow \psi_{(a)\sigma}^{R}+\Delta
\psi_{(a)\sigma}^{R}=Z^{1/2}\psi_{(a)\sigma}^{R}, \\
g_{i}&&\rightarrow Z^{-2}\left(g_{iR}+\Delta g_{iR}\right), \hspace
{0.5cm}(i=1,2),
\end{eqnarray}

\noindent where $\psi_{(a)\sigma}^{R}$ and $g_{iR}$ are,
respectively, the renormalized fields and couplings, whereas $\Delta
\psi_{(a)\sigma}^{R}$ and $\Delta g_{iR}$ are the corresponding
counterterms. Besides, $Z$ is the quasiparticle weight and measures
the coherence of the quasiparticle picture in the effective
(renormalized) description. As we explained in our previous paper
\cite{Freire1}, this parameter is related to the self-energy
effects, and as such is essential for a consistent two-loop
renormalization group analysis of the model.

\begin{figure}[t]
  \includegraphics[width=2.4in]{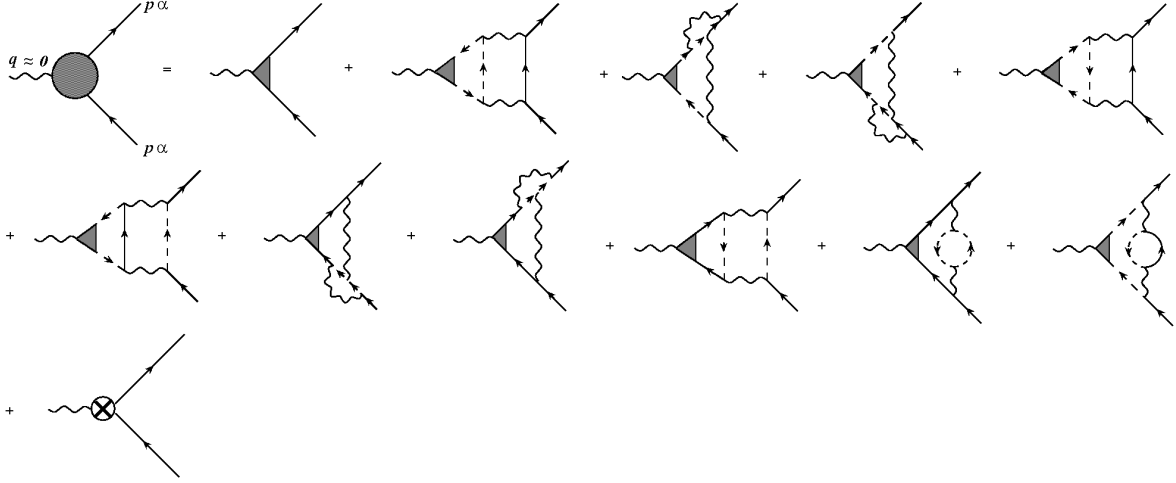}\\
  \caption{The Feynman diagrams for the uniform response function $\Gamma^{(2,1)}(p,q\approx0)$ up to two-loop order. The triangular vertices stand for
  the renormalized response function, whereas the diagram with a cross represents the counterterm.}\label{fig2}
\end{figure}

Nevertheless, there are still many ways to define the appropriate
counterterms using the above prescription. In order to solve this
ambiguity, we must establish a renormalization condition for the
observable parameters. We use here the one-particle irreducible
four-point and two-point functions (for standard definitions see,
for instance, Ref. \cite{Peskin}). Therefore, we have

\begin{equation}
\Gamma^{(4)}(p_{01}+p_{02}=p_{03}-p_{01}=\omega,
p_{1}=p_{3}=k_{F},p_{2}=p_{4}=-k_{F})=-ig_{1R}(\omega),
\end{equation}

\begin{equation}
\Gamma^{(4)}(p_{01}+p_{02}=p_{03}-p_{01}=\omega,
p_{1}=p_{4}=k_{F},p_{2}=p_{3}=-k_{F})=-ig_{2R}(\omega),
\end{equation}
\noindent and

\begin{equation}
\Gamma^{(2)}(k_{0}=\omega, k=k_{F})=\omega,
\end{equation}

\noindent where the subscript ``0'' just means that we are dealing
with the energy component associated with the external legs.
Besides, $\omega$ is an energy scale which denotes the proximity of
the renormalized theory to the FS. In this way, to flow towards the
FS, we let $\omega\rightarrow 0$.

To compute the RG flow of the renormalized couplings, one needs to
consider that the original theory (i.e. the bare theory) does not
know anything about the extrinsic scale $\omega$. In other words,
the bare couplings are independent of $\omega$. Thus, using the RG
condition $\omega d g_{i}/d\omega=0$, we obtain

\vspace {-0.3cm}

\begin{eqnarray}
\omega \frac{d g_{1R}}{d\omega}&=&\frac{g_{1R}^2}{\pi v_{F}}+\frac{g_{1R}^3}{2\pi^{2} v_{F}^{2}} \\
\omega \frac{d g_{2R}}{d\omega}&=&\frac{g_{1R}^2}{2\pi
v_{F}}+\frac{g_{1R}^3}{4\pi^{2} v_{F}^{2}}.
\end{eqnarray}

\noindent In these equations, we note that there are two
possibilities of flow as one approaches the FS. First, for models
defined with an initially repulsive backscattering coupling (i.e.
$g_{1}>0$), the low-energy effective theory naturally scales to a
line of fixed points given by $g_{1R}^{*}=0$ and $g_{2R}^{*}\neq0$.
Therefore, these systems are in the same universality class of the
so-called Tomonaga-Luttinger model where backscattering processes
are neglected altogether. For this reason, they are often referred
to as Luttinger liquids. This is the case for the 1D HM, in which we
are mainly interested here since it will provide a testing ground
for our later analysis of the 2D HM. We note, in passing, that there
is another possibility of flow which happens when the initial model
includes a backscattering coupling which is attractive (i.e.
$g_{1}<0$). In this case, the backscattering coupling becomes
relevant in the RG sense, and the resulting theory is in the same
universality class of another well-known model in 1D, which goes
under the name of Luther-Emery model. For simplicity, we will only
deal with the first case scenario in this paper.

We next derive a flow equation to analyze the quasiparticle weight
$Z$ of the Luttinger liquid universality class as one approach the
FS. Calculating the Feynman diagrams of Fig. 2, and using the
renormalization condition of Eq. (7), we obtain

\begin{equation}
\omega\frac{d \ln
Z}{d\omega}=\frac{1}{4\pi^{2}v_{F}^{2}}(g_{1R}^{2}+g_{2R}^{2}-g_{1R}g_{2R})\equiv\gamma
\end{equation}

\noindent where $\gamma$ is the so-called anomalous dimension. In
the vicinity of the fixed point $g_{1R}^{*}=0$ and
$g_{2R}^{*}\neq0$, the quasiparticle weight scales as
$Z\sim(\omega/\Omega)^{\gamma^{*}}$. Since $\gamma^{*}>0$, Z becomes
suppressed at the FS if we let $\omega\rightarrow 0$ . This result
asserts that there are no coherent low-lying quasiparticles in these
systems.

In order to obtain the uniform susceptibilities of the system, we
must first calculate the linear response function due to an
infinitesimal uniform external field, which couples with the
occupation number operator. Thus, we add to the Lagrangian the new
term

\begin{equation}
-h_{external}\sum_{p,a=\pm}\mathcal{T}_{\alpha}\psi_{(a)\alpha}^{
\dagger}(p,t)\psi_{(a)\alpha}(p,t),
\end{equation}

\noindent where, again, everything is written in terms of the bare
parameters of the model. This will generate an additional vertex
(the one-particle irreducible uniform response function
$\Gamma^{(2,1)}(p,q\approx0)$), which will in turn contain new IR
divergent diagrams but only at two-loop order or beyond in
perturbation theory. As we will point out later in the paper, at
one-loop order, there are no singular diagrams present in this
calculation. Therefore, these contributions do not need to be
regularized and, as a result, do not enter in the RG flow equations.
For this reason, they are usually dealt with by other means such as
the RPA approximation. On the other hand, at two-loop order, one can
verify the existence of several divergent diagrams in the
calculation of the uniform response functions (see Fig. 3). This
important result will, in fact, allow a direct computation of their
associated RG flow equations, which will eventually describe how
these parameters vary as we take the physical system towards the FS
of the system.

Analogously, we rewrite the bare uniform response function
$\mathcal{T}_{\alpha}$ in terms of its renormalized counterpart
$\mathcal{T}^{R}_{\alpha}$ and the appropriate counterterm $\Delta
\mathcal{T}^{R}_{\alpha}$ in the following way

\begin{equation}
\mathcal{T}_{\alpha}=Z^{-1}\left[\mathcal{T}^{R}_{\alpha}+\Delta
\mathcal{T}^{R}_{\alpha}\right].
\end{equation}

\noindent As we pointed out above, one still needs to establish how
the experimentally observable response are to be defined. We choose
the canonical renormalization condition, i.e.,
$\Gamma^{(2,1)}(p_{0}=\omega,p=k_{F},q\approx0)=-i\mathcal{T}^{R}_{\alpha}(\omega)$,
where the scale $\omega$ plays the same role explained earlier.

Now, we can define the two different types of uniform response
functions, which simply arise from a symmetrization of the object
$\mathcal{T}^{R}_{\alpha}$ with respect to the spin projection
$\alpha$, namely

\vspace {-0.3cm}

\begin{eqnarray}
\mathcal{T}^{R}_{Charge}(\omega)&=&\mathcal{T}^{R}_{\uparrow}
(\omega)+\mathcal{T}^{R}_{\downarrow}(\omega),\\
\mathcal{T}^{R}_{Spin}(\omega)&=&\mathcal{T}^{R}_{\uparrow}
(\omega)-\mathcal{T}^{R}_{\downarrow}(\omega),
\end{eqnarray}

\noindent where $\mathcal{T}^{R}_{Charge}$ and
$\mathcal{T}^{R}_{Spin}$ are the uniform response functions in the
charge and spin channels, respectively. To derive the corresponding
flow equations, we recall that $\omega d
\mathcal{T}_{\alpha}/d\omega=0$. Thus

\vspace {-0.4cm}

\begin{eqnarray}
\omega\frac{d}{d\omega}\mathcal{T}^{R}_{Charge}&=&
\left(\frac{g_{1R}g_{2R}}{4\pi^{2}v_{F}^{2}}\right)\mathcal{T}^{R}_{Charge}\label{unifsusc1}\\
\omega\frac{d}{d\omega}\mathcal{T}^{R}_{Spin}&=&
\left(\frac{g_{1R}g_{2R}}{4\pi^{2}v_{F}^{2}}\right)\mathcal{T}^{R}_{Spin},\label{unifsusc2}
\end{eqnarray}

\noindent We observe here that the RG flow equations associated with
the charge and spin uniform susceptibilities are identical in form.
This has a simple interpretation, i.e, there is no separation of the
charge and spin degrees of freedom in our current model. In other
words, the holons and the spinons elementary excitations known to
exist in these systems have the same velocity of propagation here.
This result can be traced back to our initial Lagrangian, where we
left out from our analysis the so-called $g_{4}$ processes
associated with forward scattering processes between particles from
the same branch. These interaction processes are interesting in
their own right, but they are widely recognized to be extremely
difficult to be incorporated consistently in a full RG scheme
\cite{Solyom}. Nevertheless, as we will see next, it is still
possible to extract important information from our model concerning
the low-energy properties of the Luttinger liquid universality
class.

As we saw earlier, the Luttinger liquid fixed point is defined by
the Tomonaga-Luttinger model, i.e., $g_{1R}^{*}=0$ and
$g_{2R}^{*}\neq0$. Due to the limitations associated with a two-loop
order truncation in the present RG scheme, such a fixed point is
reached only logarithmically in the RG flow. On the other hand, it
can be proven by other means, i.e. by using the so-called Ward
identities associated with an exact conservation of left and right
particles, that this fixed point indeed holds to all orders of
perturbation theory \cite{Benfatto}. For this reason, this result is
fundamentally nonperturbative in character. Thus, in order to
properly take the physical system towards the Luttinger liquid
regime in our approach, we must set from the start the
backscattering coupling $g_{1R}$ equal to zero in Eqs.
(\ref{unifsusc1}) and (\ref{unifsusc2}). Therefore, we obtain:

\vspace {-0.4cm}

\begin{eqnarray}
\omega\frac{d}{d\omega}\mathcal{T}^{R}_{Charge}&=&0\\
\omega\frac{d}{d\omega}\mathcal{T}^{R}_{Spin}&=&0,
\end{eqnarray}

\noindent This means that both uniform response functions satisfy
trivially the RG invariance condition at the Luttinger liquid fixed
point. In addition to that, the corresponding uniform
susceptibilities (see also Fig. 4) are given by

\begin{equation}
\chi_{Charge(Spin)}^{R}(\omega)=\frac{1}{2\pi
v_{F}}\left[\mathcal{T}^{R}_{Charge(Spin)}(\omega)\right]^{2}.
\end{equation}

\noindent Thus, since the uniform response functions are RG
invariant quantities at the Luttinger liquid fixed point, so will be
the corresponding uniform susceptibilities of the system in the same
regime. This is related to the fact that the low-energy effective
theory indeed contains both gapless charge and spin collective
excitations, which are indicative of a metallic ground state. In
this way, we can conclude that the two-loop RG approach is able to
reproduce, quite simply, some very important aspects of the
so-called Luttinger liquids, namely, that its fixed point
corresponds to a metallic state with no fermionic quasiparticles
excitations, and the resulting dynamics being dominated exclusively
by bosonic charge and spin collective excitations
\cite{Solyom,Haldane,Voit}.

\begin{figure}[t]
  \centering
  \includegraphics[width=2.8in]{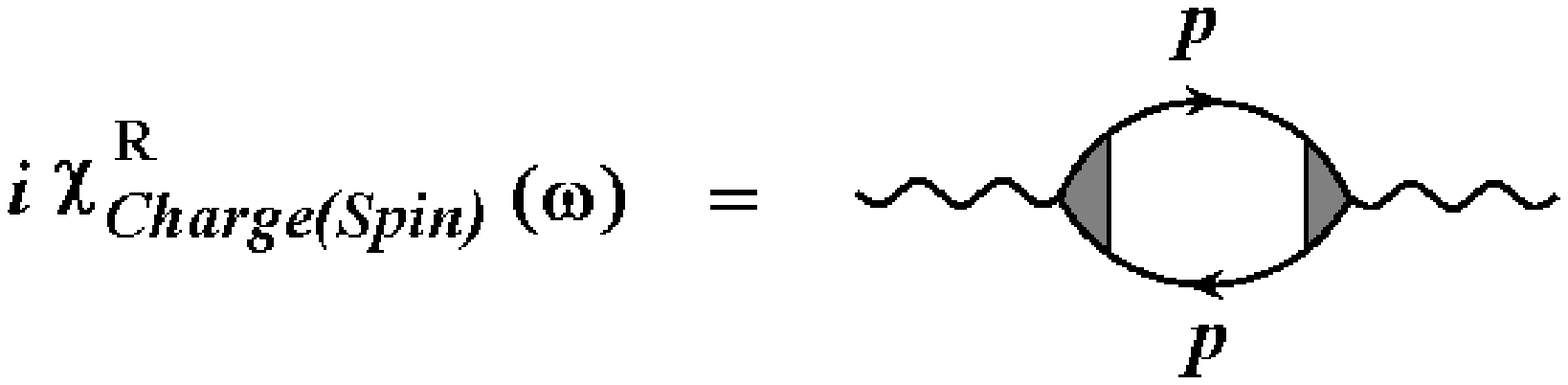}\\
  \caption{The Feynman diagram associated with the renormalized uniform charge and spin susceptibilities.}\label{fig3}
\end{figure}

\vspace {0.5cm}

\section{The 2D repulsive Hubbard model}

\vspace {0.5cm}

\subsection{The two-loop RG approach}

\vspace {0.5cm}

Our two-loop RG calculation worked quite well in reproducing some
low-energy properties of the 1D HM such as the gapless nature of the
resulting in both charge and spin excitation spectra. This result
encourages us to try to apply the same approach for problems, in
which there are no known exact solutions to this present date. This
is what happens very generally in strongly correlated 2D models, due
to the fact that well-known analytical techniques such as the
bosonization and the Bethe ansatz are not easily extendable to such
higher-dimensional systems. Here, we discuss the application of the
two-loop RG method developed earlier to the 2D repulsive HM slightly
away from half-filling.

In 2D, the repulsive ($U>0$) Hubbard Hamiltonian defined on a square
lattice is given in momentum space by

\begin{equation}
H=\sum_{\mathbf{k},\sigma}\xi_{\mathbf{k}}\psi^{\dagger}_{\mathbf{k}\sigma}\psi_{\mathbf{k}\sigma}+\left(\frac{U}{N_{sites}}\right)
\sum_{\mathbf{p},\mathbf{k},\mathbf{q}}\psi^{\dagger}_{\mathbf{p}+\mathbf{k}-\mathbf{q}\uparrow}\psi^{\dagger}_{\mathbf{q}\downarrow}
\psi_{\mathbf{k}\downarrow}\psi_{\mathbf{p}\uparrow},
\end{equation}

\noindent where the energy dispersion is given by
$\xi_{\mathbf{k}}=-2t\left[\cos(k_{x}a)+\cos(k_{y}a)\right]-\mu$,
and $\psi^{\dagger}_{\mathbf{k}\sigma}$ and
$\psi_{\mathbf{k}\sigma}$ are the creation and annihilation
operators of electrons with momentum $\mathbf{k}=(k_{x},k_{y})$ and
spin projection ${\sigma}=\uparrow,\downarrow$. The other
definitions follow those of the 1D model, i.e., $\mu$ stands for the
chemical potential, $N_{sites}$ is the number of total sites, and
$a$ is the spacing of the 2D lattice. A very important difference
here is the width of the noninteracting band, which is now given by
$W=8t$.

\begin{figure}[t]
  \centering
  \includegraphics[width=4.3in]{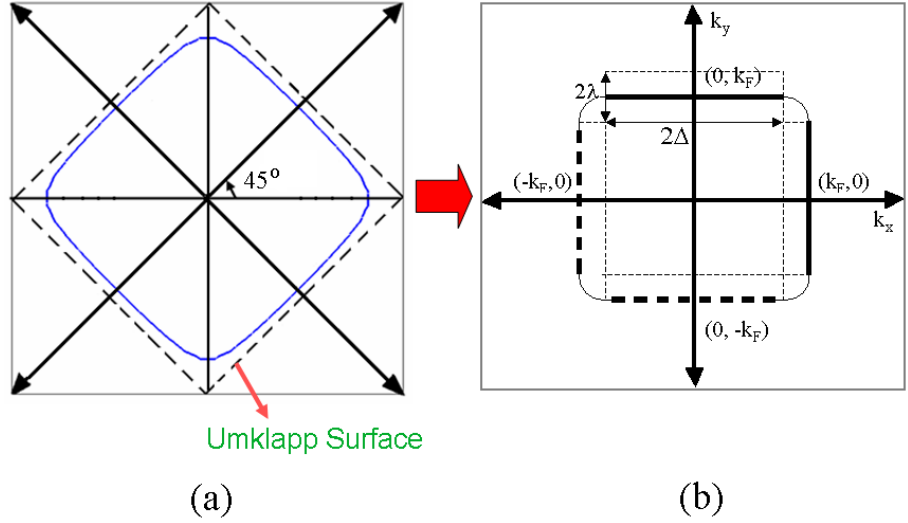}\\
  \caption{(a) The half-filled Fermi surface (FS) of the 2D Hubbard model (dashed line), the lightly hole-doped FS (solid line), and
  (b) the latter FS after a rotation of the axes by $45^{o}$ degrees.}\label{fig1}
\end{figure}

The electron band filling of the system is controlled by the ratio
$\mu/t$. When $\mu/t=0$ the model is exactly at half-filling
condition. As we start doping it with holes, $\mu/t$ takes slightly
negative values. As we can see from Fig. 5 (a), the FS of such
lightly doped system is nearly flat, and contains no van Hove
singularities in it. That FS will be our starting point for the
two-loop RG approach. However, this 2D problem has additional
complications compared to the 1D case, since we need now two momenta
$k_{x}$ and $k_{y}$ to prescribe a renormalization procedure towards
the FS. To simplify the calculations, we can perform a simple
rotation of the axes by $45^{o}$ degrees, and change these two
components by a component parallel ($k_{\parallel}$) and another
perpendicular ($k_{\perp}$) to the FS. As a result of this, the 2D
problem will bear a more transparent resemblance to the 1D case (see
Fig. 5 (b)), and we can use the machinery developed in the previous
section with some minor adaptations. To emphasize further the
similarities of both problems, we divide the 2D FS into four
different regions (two sets of solid and dashed line patches). Here,
we restrict the momenta at the FS to the flat parts only.
Analogously to the 1D problem, the interaction processes connecting
parallel patches of the FS are always logarithmically IR divergent
due to quantum fluctuations. In contrast, those connecting
perpendicular patches always remain finite, and do not contribute to
the RG flow equations in our approach. For convenience, we restrict
ourselves to one-electron states labeled by the momenta
$k_{\parallel}=k_{x}$ and $k_{\perp}=k_{y}$ associated with one of
the two sets of perpendicular patches. The momenta parallel to the
FS are restricted to the interval $-\Delta\leq k_{
\parallel}\leq\Delta$, with $2\Delta$ being
essentially the size of the flat patches. The energy dispersion of
the single-particle states is given by
$\xi_{\mathbf{k}}=v_{F}\left(\left|k_{\perp}\right|-k_{F}\right)$,
and depends only on the momenta perpendicular to the FS. We define a
label $a=\pm$ which refers to the flat sectors at $k_{\perp}=\pm
k_{F}$, respectively. In addition, we take
$k_{F}-\lambda\leq\left|k_{\perp}\right|\leq k_{F}+\lambda$, where
$\lambda$ is the fixed ultraviolet (UV) microscopic momentum cut-off
with $\Omega=2v_{F}\lambda$.

In the same spirit as before, we now write down the Lagrangian of
the 2D HM as

\begin{eqnarray}
&&L=\sum_{\textbf{k},\sigma,a=\pm}\psi_{(a)\sigma}^{\dagger}\left(\textbf{k},t\right)
\left[i\partial_{t}-v_{F}\left(\left|k_{\perp}\right|-k_{F}\right)\right]\psi_{(a)\sigma}
\left(\textbf{k},t\right)\nonumber \\
&&-\frac{1}{V}\sum_{\textbf{p},\textbf{q},\textbf{k}}\sum_{\alpha,\beta,\gamma,\delta}
\left[g_{2}\delta_{\alpha\delta}\delta_{\beta\gamma}-
g_{1}\delta_{\alpha\gamma}\delta_{\beta\delta}\right]\nonumber
\psi_{\left(+\right)\delta}^{\dagger}\left(\textbf{p}+\textbf{q}-\textbf{k},t\right)
\psi_{\left(-\right)\gamma}^{\dagger}\left(\textbf{k},t\right)
\psi_{\left(-\right)\beta}\left(\textbf{q},t\right)\psi_{\left(+\right)\alpha}\left(\textbf{p},t\right),
\\ \label{lagrangian}
\end{eqnarray}

\noindent where all the definitions follow the 1D HM case with
$\psi^{\dagger}_{(\pm)}$ and $\psi_{(\pm)}$ being fermionic fields
associated to electrons located at the $\pm$ patches and the
summation over momenta being appropriately understood as
$\sum_{\mathbf{p}}=V/(2\pi)^{2}\int d^{2}\mathbf{p}$ in the
thermodynamic limit. Since this should represent the 2D HM, the
Lagrangian is also written in a manifestly $SU(2)$ invariant form
and the couplings are initially defined as
$g_{1}=g_{2}=(V/4N_{sites})U$. In addition, we again neglect Umklapp
processes here. This choice is related to the fact that we are
mainly interested in dealing with the HM slightly away from
half-filling and, therefore, our FS does not intersect the so-called
Umklapp surface at any point (see again Fig. 5 (a)).

Apart from its physical importance especially regarding the possible
explanation of high-Tc superconductivity, this 2D model is also very
interesting from an exclusive RG point of view. In fact, when one
deals with this FS model, the counterterms needed to renormalize the
theory turn out to be continuous functions of the momenta parallel
to the FS, rather than being simply infinite shifts. As a result,
the couplings initially defined as constants in the 2D Hubbard
Lagrangian acquire a momentum dependence and become renormalized
functions of the three parallel momenta when one takes the system
towards the FS. Thus, the new prescriptions become

\begin{eqnarray}
\psi_{(a)\sigma}&&\rightarrow \psi_{(a)\sigma}^{R}+\Delta
\psi_{(a)\sigma}^{R}(k_{\parallel})=Z^{1/2}(k_{\parallel})\psi_{(a)\sigma}^{R}, \\
g_{i}&&\rightarrow\prod_{i=1}^{4}Z^{-1/2}(p_{i\parallel})\left[g_{iR}(p_{1\parallel},p_{
2\parallel},p_{3\parallel})+\Delta g_{iR}(p_{1\parallel},p_{
2\parallel},p_{3\parallel})\right].\label{coup}
\end{eqnarray}

\noindent where
$p_{4\parallel}=p_{1\parallel}+p_{2\parallel}-p_{3\parallel}$. As
before, there are still several ways to define the appropriate
counterterms using the above prescription. In order to solve this,
we establish renormalization conditions for the one-particle
irreducible four-point functions as follows \cite{Freire1}

\begin{equation}
\Gamma^{(4)}(p_{1\parallel},p_{
2\parallel},p_{3\parallel},p_{01}+p_{02}=p_{03}-p_{01}=\omega,
p_{1}=p_{3}=-p_{2}=-p_{4}=k_{F})=-ig_{1R}(p_{1\parallel},p_{
2\parallel},p_{3\parallel};\omega),
\end{equation}

\begin{equation}
\Gamma^{(4)}(p_{1\parallel},p_{
2\parallel},p_{3\parallel},p_{01}+p_{02}=p_{03}-p_{01}=\omega,
p_{1}=p_{4}=-p_{2}=-p_{3}=k_{F})=-ig_{2R}(p_{1\parallel},p_{
2\parallel},p_{3\parallel};\omega),
\end{equation}

\noindent and also for the one-particle irreducible two-point
function as

\begin{equation}
\Gamma^{(2)}(k_{0}=\omega, k=k_{F},k_{\parallel})=\omega,
\end{equation}

\noindent where $\omega$ is again the energy scale which denotes the
proximity of the renormalized theory to the FS.

Following the same procedure explained before, the corresponding RG
flow equations for the couplings and quasiparticle weight of this 2D
problem become coupled integro-differential equations, which are
impossible to solve analytically. Therefore, one must solve them
using a numerical approach. We choose here the standard fourth-order
Runge-Kutta method. We refer the reader to our paper \cite{Freire1}
for more details.

When one considers moderate initial interactions, the corresponding
quasiparticle weight $Z$ becomes suppressed as one approaches the
FS. This is a strong indicative that the resulting low-energy
effective theory does not contain low-lying electronic
quasiparticles in the system. In addition to this, instead of the
renormalized coupling functions diverging as happens in well-known
one-loop RG flows, they flow, for some particular choices of
momenta, to strong coupling plateau values. This result means that
those systems belong to the same universality class of a strongly
coupled theory, whose detailed information must be determined from a
consistent RG calculation of other quantities such as the uniform
susceptibilities as we have done explicitly for the 1D HM case. In
fact, we will see here that the suppression of the quasiparticle
weight of the system will produce interesting effects in the RG
flows of those quantities \cite{Freire2}.

First, we must calculate the linear response due to an infinitesimal
external field perturbation as before. Again, we add to the
Lagrangian the following term

\begin{equation}
-h_{external}\sum_{\mathbf{p},a=\pm}\mathcal{T}_{\alpha}(\mathbf{p})\psi_{(a)\alpha}^{B
\dagger}\left(\mathbf{p}\right)\psi^{B}_{(a)\alpha}\left(\mathbf{p}\right),
\end{equation}

\noindent This will generate an additional vertex (the one-particle
irreducible uniform response function
$\Gamma^{(2,1)}(\mathbf{p},\mathbf{q}\approx0)$), which will in turn
be afflicted by divergences given by the Feynman diagrams of Fig. 3
as in the 1D case. Therefore, we rewrite the bare quantity
$\mathcal{T}_{\alpha}$ in terms of its renormalized parameter
$\mathcal{T}_{\alpha}$ and an appropriate counterterm $\Delta
\mathcal{T}_{\alpha}$ as follows

\begin{equation}
\mathcal{T}_{\alpha}(p_{\parallel})=Z^{-1}(p_{\parallel})\left[\mathcal{T}^{R}_{\alpha}(p_{\parallel})+\Delta
\mathcal{T}^{R}_{\alpha}(p_{\parallel})\right].
\end{equation}

\noindent where the Z function is given by \cite{Freire1}

\begin{equation}
Z(p_{\parallel})=1-\frac{1}{32\pi^{4}v_{F}^{2}}\int
dk_{\parallel}dq_{\parallel}[2g_{2R}\circ g_{2R}+ 2g_{1R}\circ
g_{1R}-g_{1R}\circ g_{2R}-g_{2R}\circ
g_{1R}]\ln\left(\frac{\Omega}{\omega}\right).
\end{equation}

\noindent where the sign $\circ$ means that we are omitting the
momentum dependence of the renormalized coupling functions. As was
already mentioned, $\Delta \mathcal{T}_{\alpha}$ must absorb the
divergences generated by the Feynman diagrams of Fig. 3. However, we
may still define that counterterm in several ways. To solve this
ambiguity, we make a prescription establishing precisely that the
$\mathcal{T}^{R}_{\alpha}$ is the experimentally observable
response, i.e.,
$\Gamma^{(2,1)}(p_{\parallel},p_{0}=\omega,p_{\perp}=k_{F};\textbf{q}\approx0)=-i\mathcal{T}^{R}_{\alpha}(p_{\parallel},\omega)$,
where $\omega$ is the RG energy scale parameter.

Now we turn to the definition of the two different types of uniform
response functions, which arise from the symmetrization with respect
to the spin projection $\alpha$, i.e.
\begin{eqnarray}
\mathcal{T}^{R}_{Charge}(p_{\parallel},\omega)&=&\mathcal{T}^{R}_{\uparrow}(p_{\parallel},\omega)+\mathcal{T}^{R}_{\downarrow}(p_{\parallel},\omega),\\
\mathcal{T}^{R}_{Spin}(p_{\parallel},\omega)&=&\mathcal{T}^{R}_{\uparrow}(p_{\parallel},\omega)-\mathcal{T}^{R}_{\downarrow}(p_{\parallel},\omega),
\end{eqnarray}

\noindent where $\mathcal{T}^{R}_{Charge}$ and
$\mathcal{T}^{R}_{Spin}$ are the uniform response functions
associated with the charge and spin channels, respectively. To
compute the RG flow equations of these response functions, one needs
to recall that the bare parameters are independent of the
renormalization scale $\omega$. Thus, using the RG condition $\omega
d\mathcal{T}_{\alpha}/d\omega=0$, we obtain \cite{Freire2}

\begin{eqnarray}
&&\omega\frac{d}{d\omega}\mathcal{T}^{R}_{Charge}(p_{\parallel})=\frac{1}{32\pi^{4}v_{F}^{2}}
\bigg(\int dk_{\parallel}dq_{\parallel}[2g_{2R}\circ
g_{2R}+2g_{1R}\circ g_{1R}+g_{1R}\circ g_{2R} +g_{2R}\circ
g_{1R}\nonumber
\\&&+g_{1R}\circ g_{2R}+g_{2R}\circ g_{1R}-2g_{2R}\circ
g_{2R}-2g_{1R}\circ g_{1R} -2g_{2R}\circ g_{2R}-2g_{1R}\circ
g_{1R}]\mathcal{T}^{R}_{Charge}(q_{\parallel})\nonumber\\&&+ \int
dk_{\parallel}dq_{\parallel}[2g_{2R}\circ g_{2R}+
2g_{1R}\circ g_{1R}-g_{1R}\circ g_{2R}-g_{2R}\circ g_{1R}]\mathcal{T}^{R}_{Charge}(p_{\parallel})\bigg),\\
\nonumber \\ \nonumber \\
&&\omega\frac{d}{d\omega}\mathcal{T}^{R}_{Spin}(p_{\parallel})=\frac{1}{32\pi^{4}v_{F}^{2}}
\bigg(\int dk_{\parallel}dq_{\parallel}[g_{1R}\circ g_{2R}
+g_{2R}\circ g_{1R}+g_{1R}\circ g_{2R}+g_{2R}\circ
g_{1R}\nonumber\\&&-2g_{2R}\circ g_{2R}-2g_{1R}\circ g_{1R}
-2g_{2R}\circ g_{2R}-2g_{1R}\circ
g_{1R}]\mathcal{T}^{R}_{Spin}(q_{\parallel})+
\mathcal{T}^{R}_{Spin}(p_{\parallel})\nonumber
\\&&\times\int
dk_{\parallel}dq_{\parallel}[2g_{2R}\circ g_{2R}+ 2g_{1R}\circ
g_{1R}-g_{1R}\circ g_{2R}-g_{2R}\circ g_{1R}]\bigg),
\end{eqnarray}

\noindent where the sign $\circ$ represents again the momentum
dependence of the renormalized coupling functions which were omitted
for convenience. To find the solutions for these RG equations, we
resort to the same numerical method as described before.

Finally, once the response functions are obtained, we can calculate
the flow of the uniform charge and spin susceptibilities of the
system. They are given by the same diagram as in the 1D case (Fig.
4). Therefore, we have

\begin{equation}
\chi^{R}_{Charge(Spin)}(\omega)=\frac{1}{4\pi^{2}v_{F}}\int_{-\Delta}^{\Delta}dp_{\parallel}\left[\mathcal{T}^{R}_{Charge(Spin)}(p_{\parallel},\omega)\right]^2.\\
\end{equation}

\noindent Once more, we use the same numerical procedure to estimate
those quantities. Our results are displayed in Fig. 6. We note in
this plot that both uniform susceptibilities flow at the same rate
as we approach the FS. As a result, again, there is no spin-charge
separation effects in this model. One way to possibly circumvent
this would be to add $g_{4}$-type interaction contributions and work
hard to implement an appropriate RG scheme in this system (for this
question, see also Ref. \cite{Ferraz} for a discussion of a closely
related problem in 1D). Notice that, for moderate initial couplings
with $Z$ going to zero, the susceptibilities become strongly
suppressed in the same low-energy limit in markedly contrast to the
1D case, where they remain finite. Therefore, our result adds
further support to the interpretation that the low-energy effective
theory for those cases should be a fully gapped state in both charge
and spin excitation spectra \cite{Freire2}. Since such an effective
theory has only gapful excitations present, it cannot be related to
any spontaneously broken symmetry state. Consequently, it should
stay in a liquid phase-like state associated with a short-range
order in the system. These effects become even more transparent when
we calculate the susceptibilities associated with several order
parameters (see Ref. \cite{Correa}). There, we show that all
susceptibilities are finite as we vary the RG energy scale. Thus, it
is not possible to any given order to become long-range correlated
in this 2D system. This unusual quantum state predicted long ago by
Anderson is commonly referred to as an insulating spin liquid (ISL),
and it is widely believed to play a central role in the appropriate
explanation of the underlying mechanism of high-Tc
superconductivity.

\begin{figure}[t]
  \centering
  \includegraphics[width=3.4in]{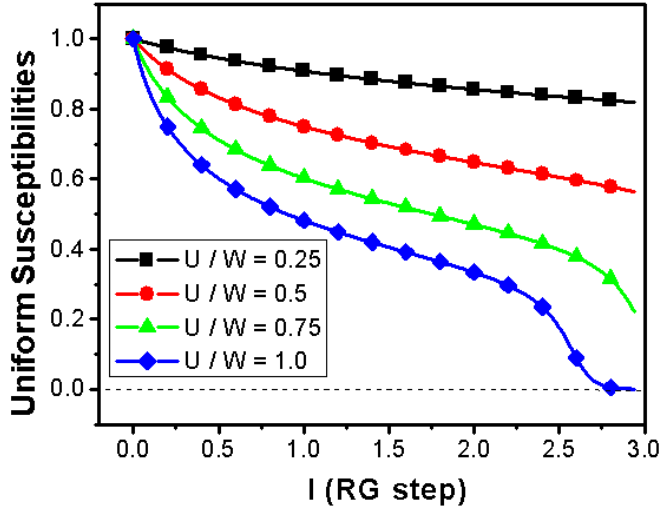}\\
  \caption{The RG flow of the uniform charge and spin susceptibilities as we increase the bare interaction strength
  given by $U=(g/\pi v_{F})t$.
  Our RG step $l$ is given by $l=\ln(\Omega/\omega)$.}\label{fig4}
\end{figure}

\vspace {0.5cm}

\subsection{The RPA approach}

\vspace {0.5cm}

We now compare the two-loop RG result with another approach for
calculating the uniform susceptibilities that appears widely in the
literature \cite{Halboth,Honerkamp}, i.e., the random phase
approximation (RPA). This method is used for estimating those
quantities at an one-loop RG level. At this order, the renormalized
couplings are known to diverge at finite energy scales, below which
the RG is simply unable to access. The approximation consists of the
assumption that the remaining information is given by the one-loop
diagrams of the uniform response functions (see Fig. 7). However,
since these diagrams are not IR divergent, it is not possible to
derive RG flow equations for those quantities at this order. The
resulting equations will be integral equations which must be solved
self-consistently in strong contrast to our previous two-loop RG
approach, which allowed us to derive proper flow equations.

Let us consider Eq. (26). If one goes only up to one-loop order, the
quasiparticle weight $Z$ must be equal to unity, since one is not
taking into account self-energy corrections in the calculations. In
this way, we have

\begin{equation}
\mathcal{T}_{\alpha}(p_{\parallel})=\mathcal{T}^{R}_{\alpha}(p_{\parallel})+\Delta
\mathcal{T}^{R}_{\alpha}(p_{\parallel}).
\end{equation}

\noindent Now, calculating the Feynman diagrams of Fig. 7 and
establishing the same renormalization condition used in the previous
section, i.e., $\Gamma^{(2,1)}(p_{\parallel},p_{0}=\omega,p_{\perp}=
k_{F};\textbf{q}\approx0)=-i\mathcal{T}^{R}_{\alpha}(p_{\parallel},\omega)$,
we obtain the following expression

\begin{equation}
\mathcal{T}^{R}_{\alpha}(p_{\parallel})=\mathcal{T}_{\alpha}(p_{\parallel})+\frac{1}{2\pi
v_{F}}\int_{-\Delta}^{\Delta} dk_{\parallel}
\left[\mathcal{T}^{R}_{\alpha}(k_{\parallel})g_{1R}(p_{\parallel},k_{\parallel},p_{\parallel})-\sum_{\sigma}
\mathcal{T}^{R}_{\sigma}(k_{\parallel})g_{2R}(p_{\parallel},k_{\parallel},k_{\parallel})\right].
\end{equation}

\begin{figure}[t]
  \centering
  \includegraphics[width=4in]{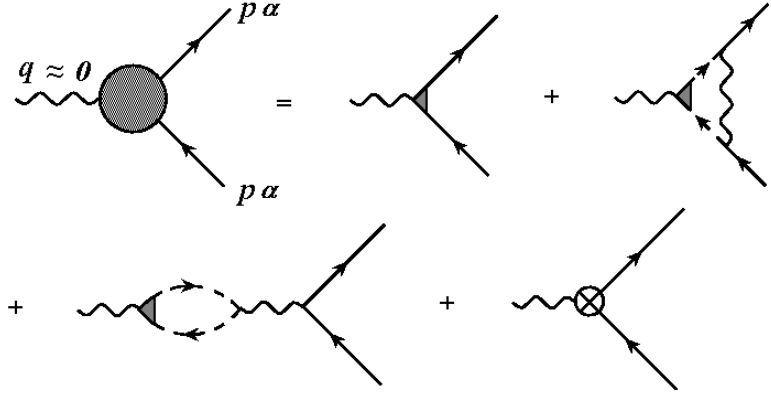}\\
  \caption{The Feynman diagrams for the calculation of the uniform response functions within the RPA approach.}\label{fig4}
\end{figure}

\noindent We note here that this equation is indeed not a RG flow
equation. Thus, symmetrizing this quantity with respect to the spin
projection $\alpha$, i.e.

\begin{eqnarray}
\mathcal{T}^{R}_{Charge}(p_{\parallel},\omega)&=&\mathcal{T}^{R}_{\uparrow}(p_{\parallel},\omega)+\mathcal{T}^{R}_{\downarrow}(p_{\parallel},\omega),\\
\mathcal{T}^{R}_{Spin}(p_{\parallel},\omega)&=&\mathcal{T}^{R}_{\uparrow}(p_{\parallel},\omega)-\mathcal{T}^{R}_{\downarrow}(p_{\parallel},\omega),
\end{eqnarray}

\noindent we obtain for the uniform charge and spin response
functions the following equations

\begin{eqnarray}
\mathcal{T}^{R}_{Charge}(p_{\parallel};\omega)&=&\mathcal{T}_{Charge}(p_{\parallel})+\frac{1}{2\pi
v_{F}}\int_{-\Delta}^{\Delta}
dk_{\parallel}\mathcal{T}^{R}_{Charge}(k_{\parallel})\nonumber\\
&&\times\left[g_{1R}(p_{\parallel},k_{\parallel},p_{\parallel};\omega)-2g_{2R}(p_{\parallel},k_{\parallel},k_{\parallel};\omega)\right],
\nonumber\\
\\\mathcal{T}^{R}_{Spin}(p_{\parallel};\omega)&=&\mathcal{T}_{Spin}(p_{\parallel})+\frac{1}{2\pi
v_{F}}\int_{-\Delta}^{\Delta}
dk_{\parallel}\mathcal{T}^{R}_{Spin}(k_{\parallel})
g_{1R}(p_{\parallel},k_{\parallel},p_{\parallel};\omega). \nonumber \\
\end{eqnarray}

\begin{figure}[t]
  \centering
  \includegraphics[width=3.4in]{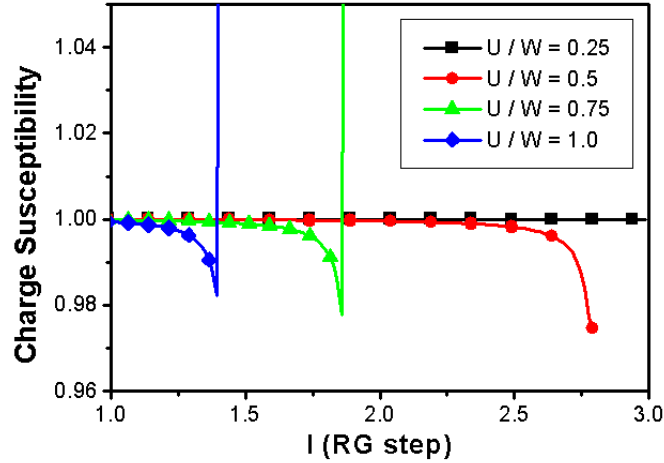}\\
  \caption{The RG flow of the uniform charge susceptibility as we increase the bare interaction strength
  given by $U=(g/\pi v_{F})t$ within the RPA approach.
  Our RG step $l$ is given by $l=\ln(\Omega/\omega)$.}\label{fig4}
\end{figure}

\noindent which now must be calculated self-consistently. Here, we
discretize the FS in exactly the same way as explained in the
previous section.

Once these quantities are estimated, we can calculate the uniform
susceptibilities in the RPA approximation, i.e.

\begin{equation}
\chi^{(RPA)}_{Charge(Spin)}(\omega)=\frac{1}{4\pi^{2}v_{F}}\int_{-\Delta}^{\Delta}dp_{\parallel}\left[\mathcal{T}^{R}_{Charge(Spin)}(p_{\parallel},\omega)\right]^2.\\
\end{equation}

\begin{figure}[t]
  \centering
  \includegraphics[width=3.4in]{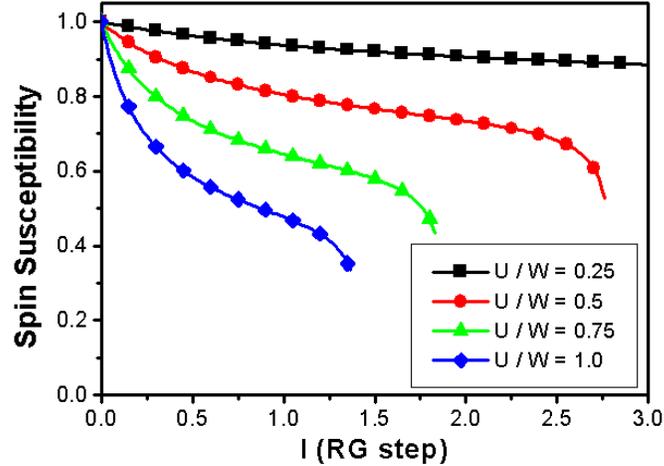}\\
  \caption{The RG flow of the uniform spin susceptibility as we increase the bare interaction strength
  given by $U=(g/\pi v_{F})t$ within the RPA approach.
  Our RG step $l$ is given by $l=\ln(\Omega/\omega)$.}\label{fig4}
\end{figure}

\noindent The results for these quantities are displayed in Figs. 8
and 9. In the charge susceptibility plot, although the initial flow
gives the impression that it becomes somewhat suppressed as we
approach the FS, we observe in fact a strong upturn of the flow
towards infinity in qualitative agreement with other numerical
results \cite{Halboth,Honerkamp}. As for the uniform spin
susceptibility, we note a tendency to flow to zero in the low-energy
limit as we increase the initial coupling $U$. Therefore, even
though one might argue that the RPA approach captures an initial
tendency of both quantities to become suppressed, this is in fact
rather approximate. As a result, one is not able to detect clearly
the formation of the ISL state in this 2D model within this
approximation.

In contrast, our two-loop RG calculation provided a more solid
evidence that the natural candidate for the low-energy effective
description of the 2D HM slightly away from half-filling is indeed
the ISL. The underlying reason for the different behavior of the two
approaches lies in the following physical principle. Quantum
fluctuation effects are well-known to be very effective in
destroying long-range correlations especially in low-dimensional
systems. In fact, the more fluctuations are taken into account, the
more likely long-range ordered states are transformed into
short-range ordered ones. This also becomes clear as a result of our
work. In accordance with this, we can infer that if we were to
consider higher-order fluctuations such as a three-loop RG
calculation or beyond, we would approach the ISL state even more
rapidly in this 2D HM system.

\vspace {0.5cm}

\section{Conclusions}

\vspace {0.5cm}

Here we investigated the renormalization of the uniform charge and
spin susceptibilities of the 2D HM slightly away from half-filling
within a two-loop RG approach. In our calculations, we took into
account simultaneously both the renormalization of the couplings and
self-energy effects. As a warm-up example, we first applied this
approach to the 1D HM away from half-filling and reproduced several
features associated with the Luttinger liquid fixed-point: the
irrelevance in the RG sense of the so-called backscattering
interaction processes, the complete absence of low-lying
quasiparticles in the vicinity of the FS, and finally the finiteness
of the both charge and spin uniform susceptibilities indicating that
the resulting state is indeed metallic.

Once the RG technique was explained in detail, we moved on to the
problem we were mostly interested in, i.e., the 2D HM for densities
slightly away from half-filling condition. We emphasized the strong
similarities to the 1D case, and adapted the technique to deal
properly with this 2D model. We pointed out that, for moderate bare
interaction $U$, the coupling functions flow to strong-coupling
plateau values for particular choices of the external momenta.
Besides, this regime is also characterized by a strongly suppressed
quasiparticle weight $Z$, which is a strong indicative of the
absence of low-lying electronic quasiparticles in the resulting
low-energy effective theory. Then, in order to extract further
information about this state, we calculated the uniform
susceptibilities associated with the charge and spin degrees of
freedom, and showed that both quantities renormalize to zero in the
low-energy limit. This result implies that both charge and spin
excitation spectra are fully gapped, and the resulting state should
be given by an insulating spin liquid.

In order to compare our result with other numerical estimates
encountered in the literature, we derived a RPA approach to
calculate the uniform susceptibilities within a one-loop RG scheme.
As a result, we reproduced the well-known one-loop results, and
showed that they were in fact not sufficient to obtain in a clear
way the ISL low-energy effective theory in this case. This should be
contrasted with the two-loop RG approach, where we do observe this
state. We explained that the physical reason for the discrepancies
of both results is that, in this particular 2D problem, one must add
more quantum fluctuations in order to effectively remove every
single instability that could potentially develop into long-range
order in the system. Therefore, if one could go beyond two-loop RG
order, one would not change this picture qualitatively, and might
well approach the ISL state even more rapidly.

\ack

\vspace {0.5cm}

This work was partially supported by the Fundação de Empreendimentos
Científicos e Tecnológicos (FINATEC) and by the Conselho Nacional de
Desenvolvimento Científico e Tecnológico (CNPq).


\begin{thebibliography}{9}
\bibitem{spinliquid} The name insulating spin liquid refers to a liquid
phase-like state, which is fully gapped in both charge and spin
excitation spectra. Therefore, it cannot be related to any broken
symmetry state and, as a consequence, should possess only
short-range order.
\end{thebibliography}
\end{document}